\documentclass[twoside]{article}
\usepackage{Proc_NTSE_16}
\usepackage{csquotes}
\usepackage{mathtools}
\usepackage{subfig}
\usepackage{graphicx}
\usepackage{dsfont}

\newcommand{\bra}[1]{\langle#1|}
\newcommand{\ket}[1]{|#1\rangle}

\def\cY{{\cal Y}}
\def\cX{{\cal X}}

\begin{document}
\thispagestyle{plain}
\publref{Raimondi\_Barbieri\_proc\_NTSE\_2016}

\begin{center}
{\Large \bf \strut
Irreducible 3-body forces \\ contributions to the self-energy 
\strut}\\
\vspace{10mm}
{\large \bf 
F. Raimondi and C. Barbieri}
\end{center}

\noindent{
\small \it Department of Physics, University of Surrey, Guildford GU2 7XH, United Kingdom}

\markboth{F. Raimondi  and  C. Barbieri}
{Including 3N forces in the many-body diagrammatic with the ADC formalism} 

\begin{abstract}
The inclusion of the three-nucleon forces (3NFs) in  \emph{ab initio} many-body approaches is 
a formidable task, due to the computational load implied by the treatment of their matrix elements. 
For this reason, practical applications have mostly been limited to contributions where 3NFs enter as effective two-nucleon interactions.
 In this contribution, we derive the algebraic diagrammatic construction (ADC) working equations for a specific Feynman diagram of the self-energy that contains a fully irreducible three-nucleon force. This diagram is expected to be the most important among those previously neglected, because it connects dominant excited intermediate state configurations. 
\\[\baselineskip] 
\end{abstract}

\section{Introduction}

The strong connection between advances in theoretical frameworks and empowering of the computational resources has emerged as one of the pillars  for the future development of the nuclear theory~\cite{Vary:2013eqa}. Different methods, such as the no-core shell model~\cite{Barrett2013131}, coupled cluster~\cite{Hagen2014}, in-medium similarity renormalization group~\cite{Hergert2016PhysRep} and self-consistent Green's function (SCGF) formalism~\cite{dickhoff2005,Dickhoff2004}, have been extended in recent years by finding efficient algorithms capable to handle the dimensionality of the nuclear many-body problem.
Most of these efforts involved novel developments of many-body formalism.
In the context of the SCGF theory applied to nuclei, different applications are currently explored. For instance, the extension of the SCFG  to encompass the concept of quasiparticle in the sense of the Bogoliubov formalism, that opens the possibility to study the open shell nuclei via the solution of the Gorkov equation~\cite{Soma2011}. The description of nuclear states in the continuum, such as electromagnetic excitations and one-nucleon elastic scattering~\cite{Idini2016}. Also the impact of three-nucleon forces (3NFs) on the mechanism of the saturation in nuclear matter~\cite{Car13} and on the correlations of  finite nuclei~\cite{Cipol13,Cipol15} has been extensively studied. 

The formalism required for the inclusion of the 3NFs in the SCGF has been laid down in Ref.~\cite{CarAr13}, where the treatment of the 3NFs in terms of effective (i.e., averaged) one- and two-nucleon forces (2NFs) is described, along with the corresponding Feynman rules for the perturbative expansion of the single-particle (s.p.) propagator.
However, the working equations for interaction-irreducible 3NFs (i.e., those diagrams that cannot simplify into effective forces) have not been investigated to date.  
 In this work we derive the working equations of one such self-energy Feynman diagram that contains a 3NF insertion not implicitly included in the effective 2NFs. Among the diagrams featuring interaction-irreducible 3NFs, we focus here on the one which is believed to be dominant, according to the energy required to excite the intermediate particle-hole configurations in the diagram. The equations are cast according to the algebraic diagrammatic construction (ADC) method, a scheme devised in quantum chemistry and applied for the first time to the perturbative expansions of the two-particle (polarization) propagator~\cite{Schirmer1982} and one-body propagator~\cite{Schirmer1983} of finite Fermi systems. The ADC allows for an efficient organization of different correlation terms in the description of the self-energy, corresponding to Feynman diagrams with different topologies such as  ladder  and ring series. Within this scheme, the nuclear Dyson equation is reformulated as an energy-independent Hermitian eigenvalue problem.  This simplifies the numerical solution, without resorting to the time-consuming algorithms that scan the entire energy spectrum in search for each pole separately. The increased dimensionality of the eigenvalue problem can be kept under control with the help of large-scale diagonalisation algorithms, such as Lanczos or Arnoldi. 
 
We present  a brief overview of the SCGF formalism in Section~\ref{intro}, covering the expressions of the Dyson equation and the irreducible self-energy. In Section~\ref{General} we review the ADC($n$) formalism up to orders $n$=2 and 3, and we outline the procedure to find the working equations for the elements for the Dyson matrix (see Eq.~\eqref{eq:Dy_residual_5} below). These working equations are given in Section~\ref{ADC2} for $n$=2 and Section~\ref{ADC3} for $n$=3. The formalism at second order is worked out in full, while at third order we limit ourselves to the set of diagrams that involve two-particle--one-hole ($2p1h$) and two-hole--one-particle ($2h1p$) intermediate configurations to illustrate the approach. In particular,  we focus on the interaction-irreducible 3NF Feynman diagram that was neglected in previous works. Finally, the conclusions are given in Section~\ref{end}.

\section{\label{intro}Basic concepts of Green's function theory}

In a microscopic approach, the description of the dynamics of the nucleus is based on a realistic interaction among the nucleons, which in principle contains different components, from the 2NF sector until the full N-body interaction. Here, we  consider up to 3NFs and start from the nuclear Hamiltonian
\begin{equation}
\label{plain_H}
\hat{H}  = \hat{T} + \hat{V} + \hat{W},
\end{equation}
with $\hat{T}$ the  kinetic energy part,  $\hat{V}$ the 2NF  and $\hat{W}$ the 3NF.

In order to treat the interaction perturbatively, we introduce the first approximation, based on the concept of the mean field felt by the nucleons as a effective external potential produced by the nuclear medium itself. Accordingly, the 
Hamiltonian is written as
\begin{eqnarray}
\label{H}
\hat{H} &=& \sum_{\alpha \beta} h^{(0)}_{\alpha \beta} \, a^\dagger_\alpha a_\beta 
 - \sum_{\alpha \beta}U_{\alpha \beta}\, a^\dagger_\alpha a_{\beta}+
\frac{1}{4} \sum_{\substack{\alpha\gamma\\\beta\delta}}V_{\alpha\gamma,\beta\delta}\, a_\alpha^\dagger a_\gamma^\dagger a_{\delta} a_{\beta}
 \nonumber \\
&&
 + \frac{1}{36}\sum_{\substack{\alpha\gamma\epsilon \\ \beta\delta\eta}} W_{\alpha\gamma\epsilon,\beta\delta\eta}\,
a_\alpha^\dagger a_\gamma^\dagger a_\epsilon^\dagger a_{\eta} a_{\delta} a_{\beta} \, ,
\end{eqnarray}
with $\hat{h}^{(0)}\equiv \hat{T} + \hat{U}$ being the mean field part, while the remaining terms give the residual interaction, which is treated in a perturbative way. The mean field part is given by the sum of the kinetic energy $T$ and the auxiliary potential $U$, defining the dynamics of the  zeroth-order propagator $g^{(0)}$ defined below, which is referred to as the mean-field reference state. 
In Eq.~(\ref{H}),  $V_{\alpha\gamma,\beta\delta}$ and $W_{\alpha \gamma\epsilon,\beta\delta\eta}$ are the \emph{antisymmetrized} matrix elements of the
2N and 3N forces respectively, with the Greek indices $\alpha$,$\beta$,$\gamma$,\ldots that label a complete set of s.p. states defining the model space used in the computation.

The peculiarity of the self-consistent Green's functions approach consists in including the solution of the dynamics of the $A$ and $A\pm 1$ nucleons systems from the start and on the same footing. This information is conveyed by the one-body propagator, or two-point Green's function. The latter is defined as the matrix element of the time-ordered  product ($\mathcal{T}$) of an annihilation and creation field operators $a(t)$ and $a^{\dagger}(t)$ with respect to the fully correlated $A$-body wave function $\ket{\Psi^{A}_0}$ in the ground state, i.e.
\begin{equation}
\label{Green}
 g_{\alpha \beta}(t-t') = - \frac{i}{\hbar}     \bra{\Psi^A_0} \mathcal{T}\left[ 	a_{\alpha}(t)  	a_{\beta}^{\dagger}(t') \right] \ket{\Psi^{A}_0} \, .
\end{equation} 
The function in Eq.~(\ref{Green}) is describing both the propagation of a particle created at time $t'$ in the quantum state $\beta$ and destroyed at a later time $t$ in the quantum state $\alpha$, and the propagation of an hole moving moving to the opposite time direction for $t'>t$. 
This is why $g(\tau)$ also takes the name of one-body propagator.

The time-coordinate representation in Eq.~(\ref{Green}) can be Fourier-transformed to the energy domain in order to obtain the 
Lehmann representation of the Green's function,
\begin{align}
 g_{\alpha \beta}(\omega) =
 \sum_n  \frac{ 
          \bra{\Psi^A_0}  	a_{\alpha}   \ket{\Psi^{A+1}_n}
          \bra{\Psi^{A+1}_n}  a^{\dagger}_{\beta}  \ket{\Psi^A_0}
              }{\hbar\omega - (E^{A+1}_n - E^A_0) + \textrm{i} \eta }  
 +\sum_k \frac{
          \bra{\Psi^A_0}      a^{\dagger}_{\beta}    \ket{\Psi^{A-1}_k}
          \bra{\Psi^{A-1}_k}  a_{\alpha}	    \ket{\Psi^A_0}
             }{\hbar\omega - (E^A_0 - E^{A-1}_k) - \textrm{i} \eta } \; ,
\label{eq:g1}
\end{align}
which contains the relevant spectroscopic informations of the $A$- and \hbox{($A\pm1$)-body} systems, contained in the transition amplitudes,
\begin{equation}
\label{tran_ampl_X}
{\cal X}^n_{\beta}\equiv\langle\Psi_n^{A+1}|a_{\beta}^{\dagger}|\Psi_0^A\rangle
\end{equation}
and
\begin{equation}
\label{tran_ampl_Y}
{\cal Y}^k_{\alpha}\equiv\langle\Psi_k^{A-1}|a_{\alpha}|\Psi_0^A\rangle \, ,
\end{equation}
which are the overlap integrals that are related to the probability of adding a particle to a orbital $\beta$ or removing it from a orbital $\alpha$ in a system with $A$ particles. In the following we will use the common notation,
\begin{equation}
\label{tran_ampl}
\mathcal{Z}_{\alpha}^{i=n,k} \equiv
\begin{cases}
({\cal X}^n_{\alpha})^*\\
{\cal Y}^k_{\alpha} \; ,
\end{cases}
\end{equation}
with the index $i$ valid for both forward-in-time (particle attachment) and backward-in-time (nucleon removal) processes. 
Note that we use $n$ to denote particle states and $k$ for hole states.
The denominators in Eq.~(\ref{eq:g1}) contain also the one-nucleon addition and removal energies
\begin{equation}
\label{e_n}
\varepsilon_n^{+}\equiv(E^{A+1}_n - E^A_0)
\end{equation}
and
\begin{equation}
\label{e_k}
\varepsilon_k^{-}\equiv(E^A_0 - E^{A-1}_k) \; ,
\end{equation}
 from which one can derive the eigenvalues corresponding to the correlated wave functions $|\Psi_n^{A\pm 1}\rangle$, once the ground state energy $E^A_0$  of  $|\Psi_0^A\rangle$ is known.  In the following, we will use the compact notations of Eqs.~(\ref{tran_ampl_X}-\ref{e_k}) to present our equations.

The s.p. Green's function~\eqref{eq:g1} is completely determined by solving the Dyson equation,
\begin{equation}
  \label{eq:Dy}
g_{\alpha\beta}(\omega)=g^{(0)}_{\alpha\beta}(\omega)+ \sum_{\gamma\delta} g^{(0)}_{\alpha\gamma}(\omega)\Sigma_{\gamma\delta}^{\star}(\omega) g_{\delta\beta}(\omega)  \; ,
\end{equation}
which is a non-linear equation for the correlated propagator, $g(\omega)$. The unperturbed propagator $g^{(0)}(\omega)$ is the propagator corresponding to the Hamiltonian $h^{(0)}$, which defines the reference state. The irreducible self-energy $\Sigma^{\star}(\omega)$ encodes the effects of the nuclear medium on the propagation and it is equivalent to the optical potential for the states in the continuum~\cite{CAPUZZI1996147,Idini2016}.

The irreducible self-energy can be separated in a term which is time-independent, $\Sigma^{\infty}$, and a energy dependent part $\widetilde{\Sigma}(\omega)$ containing contributions from the dynamical excitations given by the intermediate state configurations (ISCs) within the system:
\begin{equation}
\label{irr_SE_decomp}
\Sigma_{\alpha\beta}^{\star}(\omega) = \Sigma_{\alpha\beta}^{\infty}+ \widetilde{\Sigma}_{\alpha\beta}(\omega) \; .
\end{equation}
By inspection of the Dyson equation~\eqref{eq:Dy}, it should be clear that the self-energy contains all the effects on the propagation of the s.p. that go beyond the mean-field description: For this reason the self-energy can be regarded as an effective potential enriching the unperturbed propagator with many-body correlations  and turning it into the \enquote{dressed} propagator.
If the exact $\Sigma^{\star}(\omega)$ is know,  Eq.~\eqref{eq:Dy} yields the equivalent of the exact solution of the Schr\"odinger equation.
%

\section{\label{General} ADC formalism as matrix eigenvalue problem}

In the following we apply the algebraic diagrammatic construction to the dynamic (i.e., energy dependent) part of the irreducible self-energy of Eq.~(\ref{irr_SE_decomp}). For this purpose, we write $\widetilde{\Sigma}(\omega)$ in the most general form of its spectral representation,
\begin{eqnarray}
\label{irr_SE_Lehmann}
\begin{aligned}
\widetilde{\Sigma}_{\alpha\beta}(\omega)  =  \sum_{j j'} \textbf{M}_{\alpha j}^\dagger \Bigg[ \frac{1}{ \hbar \omega \mathds{1} - (E  \mathds{1} +  \textbf{C}) + \textrm{i} \eta\mathds{1}} \Bigg]_{\substack{j  j'}}  \textbf{M}_{j' \beta}  \\
 +  \sum_{k k'} \textbf{N}_{\alpha k} \Bigg[ \frac{1}{\hbar \omega \mathds{1} - (E \mathds{1}  +  \textbf{D}) - \textrm{i} \eta \mathds{1}} \Bigg]_{k k'}  \textbf{N}_{k' \beta }^\dagger \; .
\end{aligned}
\end{eqnarray}
At this stage, the expression in Eq.~(\ref{irr_SE_Lehmann}) is a formal decomposition defining two types of matrices with respect to the ISCs $j, j'$ ($k, k'$): the coupling matrix $\textbf{M}_{j \alpha} $ ($\textbf{N}_{\alpha k} $), and the interaction matrix $\textbf{C}_{\substack {j  j'}}$ ($\textbf{D}_{\substack{ k  k'}}$) for the forward-in-time (backward-in-time) part of the self-energy. The coupling matrices couple the initial and final s.p. states of the propagator to the ISCs, while the interaction matrices $\textbf{C}$ and $\textbf{D}$ contain the matrix elements of the residual interactions (up to 3NFs) among the ISCs themselves.
In general, ISCs  are multiparticle-multihole excitations resulting from the \hbox{same-time} propagation of fermion lines within Feynman diagrams.
For nucleon addition, with ${M+1}$ particles and $M$ holes, $(M+1)pMh$, their unperturbed energies will be:
\begin{equation}
\label{energy_E_i}
E_j = \varepsilon^{+}_{n_1} + \varepsilon^{+}_{n_2} + \cdots  ~ + \varepsilon^{+}_{n_M}  + \varepsilon^{+}_{n_{M+1}}  ~- \varepsilon^{-}_{k_1} - \varepsilon^{-}_{k_2} - \cdots ~- \varepsilon^{-}_{k_M} \; .
\end{equation}
and similarly for  nucleon removal.
In the following we will make use of the shorthand notation for the forward-in-time terms (corresponding to particle attachment)
\begin{equation}
\label{def_r}
\begin{rcases}
  r, r'&\equiv (n_1,n_2,k_3)  \\
  q, q' &\equiv  (n_1,n_2,n_3,k_4,k_5) \\
  \end{rcases}
\text{ $j$ $j'$} 
\end{equation}
and 
\begin{equation}
\label{def_s}
\begin{rcases}
  s, s' &\equiv (k_1,k_2,n_3) \\
  u, u' &\equiv (k_1,k_2,k_3,n_4,n_5)  \\
  \end{rcases}
\text{ $k$ $k'$} \; ,
\end{equation}
for the backward-in-time terms (i.e., for  particle removal). For instance,   \hbox{$j \equiv (n_1,n_2,k_3)$} in the  coupling matrix $\textbf{M}_{(n_1,n_2,k_3) \alpha } $ connects a s.p. state $\alpha$ to an intermediate state composed by a $2p1h$ configuration.
Each ISC gives a different term in Eq.~(\ref{irr_SE_Lehmann}), with the configurations $3p2h$, $4p3h$ and so on pertaining to more complicated, but also energetically less important, intermediate states. 

While the energy-dependence in the self-energy is a direct consequence of the underlying dynamics in the many-body system, it gives rise to a major computational bottleneck. In order to find all the poles of the propagator in Eq.~(\ref{eq:Dy}), one should scan the energy plane with an extremely fine mesh, therefore the direct search of the s.p. energies in this way would be costly, with the possibility to leave some solutions undetected. For this reason it is convenient to rearrange the Dyson equation in a matrix form independent of the energy.  This is achieved by introducing the eigenvector
\begin{equation}
  \label{eq:Dy_residual_6}
 \texttt{Z}^{i \dagger} \equiv \begin{pmatrix}
    \mathcal{Z}^i_{\delta}{}^{\dagger} & 
  \mathcal{W}^i_r{}^{\dagger} &
  \mathcal{W}^i_s{}^{\dagger} &
  \mathcal{W}^i_q{}^{\dagger} &
  \mathcal{W}^i_u{}^{\dagger}  &  \cdots
   \end{pmatrix} \, ,
\end{equation}
with the first component given by the transition amplitudes of Eq.~(\ref{tran_ampl}). The other components contain the information on the ISCs propagated 
through $\widetilde{\Sigma}(\omega)$ but evaluated at the specific quasiparticle energy $\varepsilon^\pm_i$ of each solution, 
\begin{equation}
  \label{eq:Dy_residual_4}
\mathcal{W}^i_j \equiv  \mathcal{W}_j(\omega) \rvert_{\hbar \omega =\varepsilon_i}  =  \sum_{j'} \Bigg[  \frac{1}{\hbar \omega \mathds{1} -(E \mathds{1} + \textbf{C} )} \Bigg]_{j j'}  \textbf{M}_{j' \delta}  \mathcal{Z}_{\delta}^i  \Bigg\rvert_{\hbar \omega =\varepsilon_i} \; ,
\end{equation}
and
%
\begin{equation}
  \label{eq:Dy_residual_4_bis}
\mathcal{W}^i_k \equiv  \mathcal{W}_k(\omega)  \rvert_{\hbar \omega =\varepsilon_i}  =  \sum_{k'}  \Bigg[  \frac{1}{\hbar \omega \mathds{1} -(E \mathds{1} + \textbf{D} )} \Bigg]_{k k'} \textbf{N}^{\dagger}_{k' \delta}
\mathcal{Z}_{\delta}^i  \Bigg\rvert_{\hbar \omega =\varepsilon_i}\; .
\end{equation}

 The task now is to diagonalize the following matrix, being equivalent to the original eigenvalue problem~\cite{Carlo2016}:
%
%
\begin{equation}
  \label{eq:Dy_residual_5}
   \resizebox{\linewidth}{!}{%
 $\epsilon_i   \texttt{Z}^{i}
\! = \!
 \begin{pmatrix}
   \varepsilon^{(0)} \! + \! \Sigma_{\alpha \delta}^\infty    &  \textbf{M}^{\dagger}_{\alpha r} & \textbf{N}_{\alpha s} &  \textbf{M}^{\dagger}_{\alpha q} & \textbf{N}_{\alpha u} & \cdots &    \\
   \\
   \textbf{M}_{r' \alpha}  &  E_r \delta_{r r'} \! + \! \textbf{C}_{\substack {r r'}} &   & \textbf{C}_{\substack {r'  q}}  &   & \cdots \\
   \\
   \textbf{N}^{\dagger}_{s' \alpha}  &  & E_s \delta_{s s'}  \! + \! \textbf{D}_{\substack {s s'}} &  & \textbf{D}_{\substack { s'  u}}  & \cdots \\
   \\
     \textbf{M}_{q' \alpha}  & \textbf{C}_{\substack {q' r}}  &  & E_q \delta_{q q'} \! + \! \textbf{C}_{\substack {q'  q}} &  &\cdots  \\
   \\
     \textbf{N}^{\dagger}_{u' \alpha}  & & \textbf{D}_{\substack {u'   s}}  &  & E_u \delta_{u u'}  \! + \! \textbf{D}_{\substack {u'  u}} & \cdots\\     
     \vdots  & \vdots & \vdots &  \ddots
\end{pmatrix} 
\texttt{Z}^{i} \, ,
$}
\end{equation}
with the normalization condition
\begin{equation}
\label{normal}
\sum_{\alpha \beta} (\mathcal{Z}_{\alpha}^i)^{\dagger} \mathcal{Z}_{\beta}^i + (\mathcal{W}^i_j)^{\dagger} \mathcal{W}^i_j +  (\mathcal{W}^i_k)^{\dagger} \mathcal{W}^i_k + \dots =1 \; .
\end{equation}

With the procedure outlined above and the introduction of the eigenvector  $\texttt{Z}^{i}$ of Eq.~(\ref{eq:Dy_residual_6}), each energy eigenvalue is now related to an eigenvector of larger dimension. 
Once Eq.~\eqref{eq:Dy_residual_5} is diagonalized, its eigenvalues and the first portions of their eigenvectors, $\mathcal{Z}^i$, yield the one-body propagator according to Eq.~\eqref{eq:g1}.
The severe growth in the dimension of the Dyson matrix can be handled by projecting the set of the energies configurations to a smaller Krylov subspace, and then a multi-pivot Lanczos-type algorithm can be applied, as illustrated in Ref.~\cite{Soma2014}.
 
The ADC is a systematic approach to find expressions for the coupling and interaction matrices appearing in Eq.~\eqref{eq:Dy_residual_5} that include  the correlations due to 2NFs,  3NFs, and so on. This is achieved by expanding Eq.~(\ref{irr_SE_Lehmann}) in powers of the residual interaction $\hat{U}$, the 2NF $\hat{V}$ and 3NF $\hat{W}$ 
and then by comparing the result with the  Goldstone-Feynman perturbative expressions for the self-energy. Formally, we have:
\begin{equation}
\label{expan_M}
\textbf{M}_{j \alpha} = \textbf{M}^{(\textrm{I})}_{j \alpha} + \textbf{M}^{(\textrm{II})}_{j \alpha} + \textbf{M}^{(\textrm{III})}_{j \alpha} + \dots \; ,
\end{equation}
where the term $\textbf{M}^{(\textrm{n})}_{j \alpha } $ is of $n^{\textrm{th}}$ order in the residual interaction.  And for the backward-in-time coupling matrices:
\begin{equation}
\label{expan_N}
\textbf{N}_{\alpha k} = \textbf{N}^{(\textrm{I})}_{\alpha k} + \textbf{N}^{(\textrm{II})}_{\alpha k} + \textbf{N}^{(\textrm{III})}_{\alpha k} + \dots \; .
\end{equation}
The  matrices $\textbf{C}$ and $\textbf{D}$ can only be at first order in the residual interaction, but they appear at the denominators in the spectral representation~\eqref{irr_SE_Lehmann}. Thus, they give rise to a geometrical series according to the identity
\begin{equation}
\label{geo_Serie}
\frac{1}{A-B} = \frac{1}{A} + \frac{1}{A} B \frac{1}{A-B}  = \frac{1}{A}  + \frac{1}{A} B \frac{1}{A} + \frac{1}{A} B \frac{1}{A}  B \frac{1}{A} +\cdots \; ,
\end{equation}
for $A=\hbar \omega - E $ and $B = \textbf{C}, \textbf{D}$.

Using  the expressions~(\ref{expan_M}-\ref{geo_Serie})  in Eq.~(\ref{irr_SE_Lehmann}) gives rise to the following expansion for the energy-dependent irreducible self-energy,
\begin{eqnarray}
\label{irr_SE_EXPA}
\widetilde{\Sigma}_{\alpha\beta}(\omega) & = & \sum_{j} \textbf{M}^{(\textrm{I}) \dagger}_{\alpha j} \left[ \frac{1}{\hbar \omega - E_j + \textrm{i} \eta} \right]  \textbf{M}^{(\textrm{I})}_{j \beta  } \nonumber \\
&& + \sum_{j} \textbf{M}^{(\textrm{II}) \dagger}_{\alpha j} \left[ \frac{1}{\hbar \omega - E_j + \textrm{i} \eta} \right] \textbf{M}^{(\textrm{I})}_{j \beta }  + \sum_{j} \textbf{M}^{(\textrm{I}) \dagger}_{\alpha j} \left[ \frac{1}{\hbar \omega - E_j + \textrm{i} \eta} \right]  \textbf{M}^{(\textrm{II})}_{j \beta } \nonumber \\
&&+  \sum_{j j'} \textbf{M}^{(\textrm{I}) \dagger}_{\alpha j} \left[ \frac{1}{\hbar \omega - E_j  + \textrm{i} \eta} \right] \textbf{C}_{\substack {j  j'}}  \left[ \frac{1}{\hbar \omega - E_{j'} + \textrm{i} \eta} \right] \textbf{M}^{(\textrm{I})}_{j' \beta } + \cdots \nonumber \\
&& +  \sum_{k} \textbf{N}^{(\textrm{I})}_{\alpha k} \left[ \frac{1}{\hbar \omega - E_k- \textrm{i} \eta} \right] \textbf{N}_{k \beta}^{(\textrm{I}) \dagger}\nonumber \\
&& +  \sum_{k} \textbf{N}^{(\textrm{II})}_{\alpha k} \left[ \frac{1}{\hbar \omega - E_k- \textrm{i} \eta} \right] \textbf{N}_{k \beta}^{(\textrm{I}) \dagger} +
 \sum_{k} \textbf{N}^{(\textrm{I})}_{\alpha k} \left[ \frac{1}{\hbar \omega - E_k- \textrm{i} \eta} \right] \textbf{N}_{k \beta }^{(\textrm{II}) \dagger} \nonumber \\
&&+  \sum_{k k'} \textbf{N}^{(\textrm{I})}_{\alpha k} \left[ \frac{1}{\hbar \omega - E_k  - \textrm{i} \eta} \right] \textbf{D}_{\substack {k k'}}  \left[ \frac{1}{\hbar \omega - E_{k'}  - \textrm{i} \eta} \right]  \textbf{N}^{(\textrm{I}) \dagger}_{k' \beta} + \cdots \; ,
\end{eqnarray}
where we show all contributions up to second and third order. The procedure is to compare term by term the formal expansion~(\ref{irr_SE_EXPA}) with the calculated  Goldstone-type diagrams. One then extracts the minimal expressions for the matrices $\textbf{M}$, $\textbf{N}$, $\textbf{C}$ and $\textbf{D}$, given in terms of the transitions amplitudes of Eqs.~(\ref{tran_ampl_X}-\ref{tran_ampl_Y}) and the quasiparticle energies of Eqs.~(\ref{e_n}-\ref{e_k}), that ensure consistency with the standard perturbative expansion up to order $n$.
The content of the ADC($n$) expansion is far from trivial when one moves from the second to the third order: In fact the structure of the third-order terms in Eq.~(\ref{irr_SE_EXPA}), as analytic functions in the energy plane,  does not match the general spectral representation of Eq.~(\ref{irr_SE_Lehmann}), which is required for the correct self-energy.
However, once $\textbf{M}$, $\textbf{N}$, $\textbf{C}$ and $\textbf{D}$ are found one can insert them in the correct analytical representation of Eqs.~\eqref{irr_SE_Lehmann}  and~\eqref{eq:Dy_residual_5}. 
As a result, the ADC($n$) approach will include selected contributions at order higher than $n$, as well as all-order
 non-perturbative resummations as shown by Eq.~\eqref{geo_Serie}.

%

\section{\label{sec:ADC(2_3)} ADC equations up to third order}
In this section we collect the building blocks of the ADC at second order and present a selected set of coupling and interaction matrices that play a dominant role at third order, in a sense to be specified in the following discussion.

All the diagrams discussed in this work are one-particle irreducible, skeleton and interaction-irreducible diagrams. When limiting oneself to only interaction-irreducible diagrams, one needs to substitute the original one- and two-nucleon residual interactions, $-\hat{U}$ an $\hat{V}$, with corresponding effective interactions, which we label respectively $\widetilde{U}$  and $\widetilde{V}$ and represent diagrammatically as wavy lines.  The latter 
 contain averaged contributions from 3NFs that account for the discarded interaction-reducible diagrams. Hence, one reduces the number of perturbative terms (i.e., diagrams) that need to be dealt with.  A detailed exposition of these aspects  and the extension of Feynman rules to the case of many-nucleon interactions is beyond the scope of the present work. The interested reader is referred to the thorough discussion in Ref.~\cite{CarAr13}.
For the present discussion, we only need to keep in mind that the  2NFs in Figs.~\ref{2ord_2B} 
are effective interactions which contain the most 'trivial' contributions of $\hat{W}$, in the sense that they do not require any extension of the formalism and computer codes previously developed for pure two-body interactions.

\subsection{\label{ADC2} ADC(2) building blocks}
\begin{figure}[t]
  \centering
    \subfloat[]{\label{2ord_2B}\includegraphics[scale=0.55]{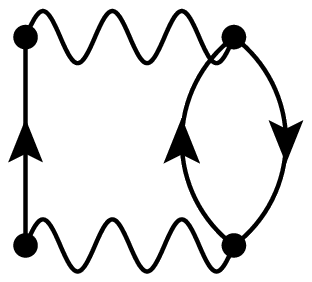}}
  \hspace{2cm}
   \subfloat[]{\label{2ord_3B}\includegraphics[scale=0.55]{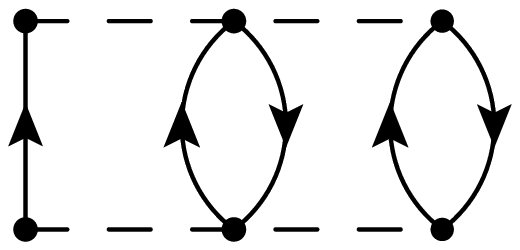}}
  \caption{ One-particle irreducible, skeleton and interaction-irreducible self-energy diagrams appearing at second order in the perturbative expansion of the self-energy. The wiggly lines represent the effective 2NF $\widetilde{V}$ containing averaged 3NF components, while the long-dashed lines represent the interaction-irreducible 3NF, $\hat{W}$.}
  \label{2ord}
\end{figure}

At second order and with 3NFs, the dynamic self-energy is composed by the two diagrams depicted in  Fig.~\ref{2ord}.   The main topological difference between them is given by the fact that Fig.~\ref{2ord_2B} propagates  $2p1h$ and $2h1p$ as intermediate states,
whereas the diagram of Fig.~\ref{2ord_3B} contains irreducible 3NFs that generate $3p2h$ and $3h2p$ ISCs. 
Since the latter are energetically less favourable, they are expected  to play a minor role at the Fermi surface and to contribute weakly to the total ground state energy. Following the same argument, we can expect that the third-order diagrams  containing $2p1h$ and $2h1p$ ISCs, discussed in Sec.~\ref{ADC3} (see Fig.~\ref{3ord_a_b_c}), are more important than the ones with $3p2h$ and $3h2p$ configurations, at the same order.

To define the ADC(2) approximation scheme, we present the explicit expressions of the coupling and interaction matrices contained in the diagrams of Fig.~\ref{2ord}. Unless otherwise stated, in this section and in the rest of the paper we adopt the Einstein's convention of summing over repeated indexes for both the model-space s.p. states ($\alpha$, $\beta$, $\dots$) and the particle and hole orbits ($n_1$, $n_2$, $\dots$, $k_1$, $k_2$, $\dots$). We also use collective indexes for ISCs according to the notation set in Eqs.~(\ref{def_r}-\ref{def_s}), where appropriate.

\begin{figure}[t]
  \centering
    \subfloat[]{\label{M_2a_nnk}\includegraphics[scale=0.45]{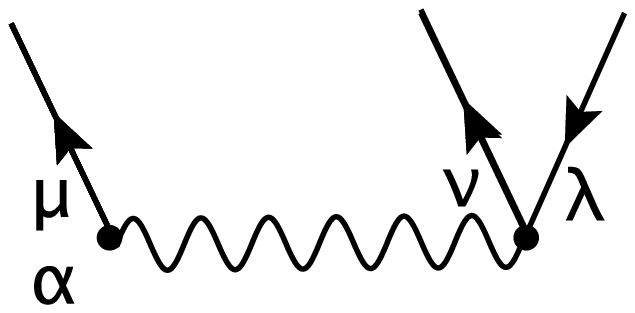}}
  \hspace{1.cm}
    \subfloat[]{\label{M_2b_nnnkk}\includegraphics[scale=0.45]{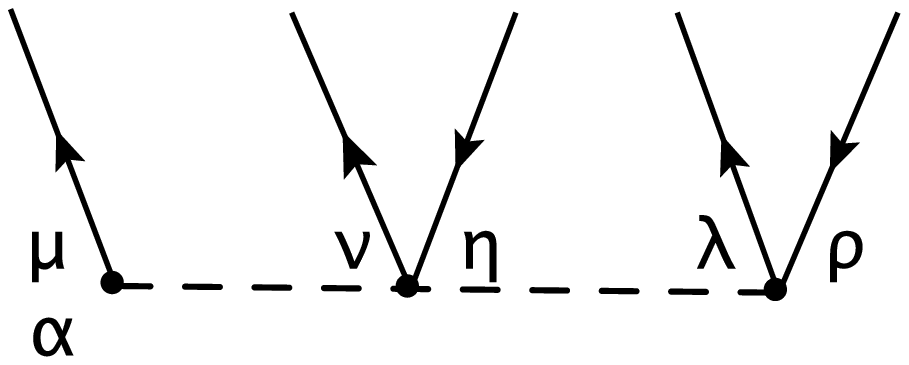}}
    \\
    \subfloat[]{\label{N_2a_kkn}\includegraphics[scale=0.45]{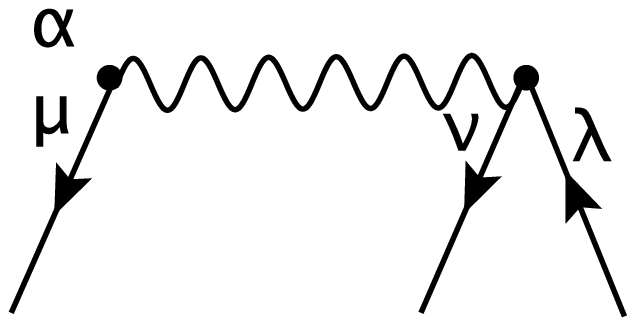}}
  \hspace{1cm}
    \subfloat[]{\label{N_2b_kkknn}\includegraphics[scale=0.45]{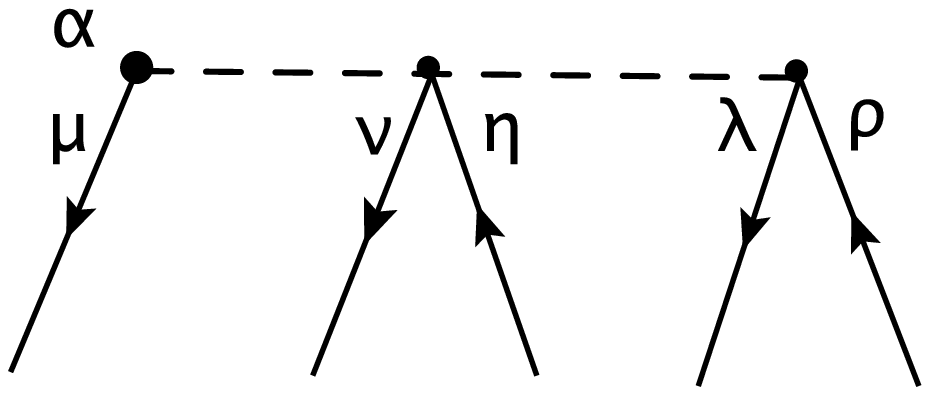}}
  \caption{ Diagrams of the ADC(2) coupling matrices containing effective 2N interaction $\widetilde{V}$ (left column) and interaction-irreducible 3NF $\hat{W}$ (right column). The coupling matrices (a) and (c) connect with $2p1h$ and $2h1p$ ISCs respectively (see Eqs.~(\ref{eq:M_2a},\ref{eq:N_2a})), while the coupling matrices (b) and (d) connect with $3p2h$ and $3h2p$ ISCs respectively (see Eqs.~(\ref{eq:M_2b},\ref{eq:N_2b})). The diagrams~\ref{M_2a_nnk} and~\ref{M_2b_nnnkk} contribute to $\textbf{M}^{(\textrm{I})}$, while ~\ref{N_2a_kkn} and~\ref{N_2b_kkknn} contribute to $\textbf{N}^{(\textrm{I})}$.}
  \label{M_N_2order}
\end{figure}

We show first the expressions  for the energy-dependent self-energy of Fig.~\ref{2ord_2B},
\begin{eqnarray}
\label{1a_expr}
 \widetilde{\Sigma}^{(1a)}_{\alpha\beta }(\omega) \! \! \! &=&
\! \! \!  \frac{1}{2}
  \widetilde{V}_{\alpha \epsilon,  \gamma \rho}
 \left(
\sum_{n_1,n_2,k_3} \frac{(\cX_{\gamma}^{n_1} \cX_{\rho}^{n_2} \cY_{\epsilon}^{k_3} )^* \cX_{\mu}^{n_1} \cX_{\nu}^{n_2}  \cY_{\lambda}^{k_3}}{\hbar \omega -\left(\varepsilon_{n_1}^{+}+\varepsilon_{n_2}^{+}-\varepsilon_{k_3}^{-} \right) + i\eta  }  \right. \nonumber \\
&&\left.\! \! \!   + \! \sum_{k_1,k_2,n_3 } \! \frac{ \cY_{\gamma}^{k_1} \cY_{\rho}^{k_2} \cX_{\epsilon}^{n_3} 
(\cY_{\mu}^{k_1} \cY_{\nu}^{k_2}  \cX_{\lambda}^{n_3})^*
 }{\hbar \omega - (\varepsilon_{k_1}^{-}+\varepsilon_{k_2}^{-}-\varepsilon_{n_3}^{+} ) - i \eta }  \right) \widetilde{V}_{\mu \nu ,  \beta \lambda} \; ,
\end{eqnarray}
and in Fig.~\ref{2ord_3B},
\begin{eqnarray}
\label{1b_expr}
 \widetilde{\Sigma}^{(1b)}_{ \alpha \beta}(\omega) \! \! \! &=&
\! \! \! \frac{1}{12} \,
  W_{\alpha \gamma \delta ,\xi \tau \sigma}
\left( \sum_{\substack{n_1,n_2,n_3 \\ k_4,k_5}}
  \frac{(\cX_{\xi}^{n_1} \cX_{\tau}^{n_2} \cX_{\sigma}^{n_3}  \cY_{\delta}^{k_4} \cY_{\gamma}^{k_5}  )^* \cX_{\mu}^{n_1}  \cX_{\nu}^{n_2} \cX_{\lambda}^{n_3} \cY_{\rho}^{k_4} \cY_{\eta}^{k_5}}{\hbar \omega -\left(\varepsilon_{n_1}^{+}+\varepsilon_{n_2}^{+}+\varepsilon_{n_3}^{+}-\varepsilon_{k_4}^{-}-\varepsilon_{k_5}^{-}  \right) + i\eta }  \right. \nonumber \\
&&\left. \!  \! + \! \sum_{\substack{k_1,k_2,k_3 \\ n_4,n_5}} \frac{
  \cY_{\xi}^{k_1} \cY_{\tau}^{k_2} \cY_{\sigma}^{k_3} \cX_{\delta}^{n_4}  \cX_{\gamma}^{n_5} (\cY_{\mu}^{k_1}  \cY_{\nu}^{k_2} \cY_{\lambda}^{k_3} \cX_{\rho}^{n_4} \cX_{\eta}^{n_5})^* }{\hbar \omega - (\varepsilon_{k_1}^{-}+\varepsilon_{k_2}^{-}+\varepsilon_{k_3}^{-}-\varepsilon_{n_4}^{+} -\varepsilon_{n_5}^{+}) - i\eta } \! \right) \! W_{\mu \nu \lambda, \beta \eta \rho} \; .
\end{eqnarray}
These expressions at second order in the Feynman-Goldstone perturbative expansion, already  match the second-order terms in the analogous expansion of the self-energy (first and fourth lines in Eq.~\eqref{irr_SE_EXPA}). Since they are already in the correct form of the spectral representation, Eq.~\eqref{irr_SE_Lehmann}, it is easy to read the  coupling
matrices at the ADC(2) level directly from them. In the first line of  Eq.~(\ref{1a_expr}) we find the coupling matrix
\begin{equation}
  \label{eq:M_2a}
\textbf{M}^{(\textrm{I-2N})}_{(n_1 n_2 k_3) \alpha }  \equiv \frac{1}{\sqrt{2}} \, \cX_{\mu}^{n_1}  \cX_{\nu}^{n_2} \cY_{\lambda}^{k_3}   \ \widetilde{V}_{\mu\nu,\alpha \lambda} \; ,
  \end{equation}
  while in the backward-in-time part (second line of Eq.~(\ref{1a_expr}))  we have
\begin{equation}
  \label{eq:N_2a}
\textbf{N}^{(\textrm{I-2N})}_{\alpha (k_1 k_2 n_3)} \equiv \frac{1}{\sqrt{2}} \, \widetilde{V}_{\alpha \lambda,\mu\nu} \ \cY_{\mu}^{k_1}  \cY_{\nu}^{k_2} \cX_{\lambda}^{n_3} \; ,
  \end{equation}
  that couples s.p. states to the $2h1p$ propagator through an effective 2NF. Both these coupling matrices can be depicted as fragments of  Goldstone diagrams, as shown in Figs.~\ref{M_2a_nnk} and~\ref{N_2a_kkn}.

  The matrix coupling  to $3p2h$ ISCs is found from Eq.~(\ref{1b_expr}) and comes from the  diagram of Fig.~\ref{2ord_3B},
  \begin{equation}
  \label{eq:M_2b}
\textbf{M}^{(\textrm{I-3N})}_{(n_1 n_2 n_3 k_4 k_5) \alpha } \equiv \frac{1}{\sqrt{12}} \,  \cX_{\mu}^{n_1} \cX_{\nu}^{n_2}  \cX_{\lambda}^{n_3}  \cY_{\rho}^{k_4} \cY_{\eta}^{k_5} \, W_{\mu\nu\lambda,\alpha\eta \rho} \; ,
  \end{equation}
    while the coupling matrix to $3h2p$ ISCs has the following expression,
    \begin{equation}
  \label{eq:N_2b}
\textbf{N}^{(\textrm{I-3N})}_{\alpha (k_1 k_2 k_3 n_4 n_5)} \equiv \frac{1}{\sqrt{12}} \, W_{\alpha\eta \rho,\mu\nu\lambda} \, \cY_{\mu}^{k_1} \cY_{\nu}^{k_2}  \cY_{\lambda}^{k_3}  \cX_{\rho}^{n_4} \cX_{\eta}^{n_5} \; .
  \end{equation}
  Their representation as fragments of Goldstone diagrams is given in Figs.~\ref{M_2b_nnnkk} and~\ref{N_2b_kkknn}, respectively.

All four expressions of Eqs.~(\ref{eq:M_2a}-\ref{eq:N_2b}) are building blocks of the ADC(2).
These complete the set of coupling matrices needed to reproduce the second order terms (first and fourth row) in Eq.~(\ref{irr_SE_EXPA}) and no interaction matrix is needed at this order.
Hence, the ADC(2)  working equations are finally summarized by Eq.~(\ref{energy_E_i}) and the following expressions:
\begin{align}
\label{ADC2_somm_I}
\textbf{M}^{(\textrm{I})}_{j \alpha}  ={}&
\left\{ \begin{array}{lcl} 
 \textbf{M}^{(\textrm{I-2N})}_{ r  \alpha}     &\qquad& \hbox{for } j= r=(n_1 n_2 k_3) \; , \\ 
 \textbf{M}^{(\textrm{I-3N})}_{ q \alpha }   && \hbox{for } j= q =(n_1 n_2 n_3 k_4 k_5) \; ,
\end{array} \right.
 \\
\label{ADC2_somm_II}\textbf{N}^{(\textrm{I})}_{\alpha k}  ={}&
\left\{ \begin{array}{lcl} 
 \textbf{N}^{(\textrm{I-2N})}_{\alpha s}     &\qquad& \hbox{for } k= s=(k_1 k_2 n_3) \; , \\ 
 \textbf{N}^{(\textrm{I-3N})}_{\alpha u}  && \hbox{for } k= u =(k_1 k_2 k_3 n_4 n_5) \; ,  
 \end{array} \right.
\\
\label{somm_C}\textbf{C}_{j j'} ={}&  0 \; , \\
\label{somm_D}\textbf{D}_{k k'}  ={}& 0 \; .
\end{align}
In ADC(2), the coupling matrices are linked directly without any intermediate interaction insertion, therefore the interaction matrices $\textbf{C}$ and $\textbf{D}$ in Eqs.~(\ref{somm_C}-\ref{somm_D}) are set to zero. As we show below, this is not  anymore true for ADC(3), where the  interaction matrices $\textbf{C}$ and $\textbf{D}$ no longer vanish and give rise to infinite (and non-perturbative) resummations of diagrams. 

\subsection{\label{ADC3} ADC(3) building blocks with $2p1h$ and $2h1p$ ISCs}

\begin{figure}[t]
  \centering
    \subfloat[]{\label{3ord_a}\includegraphics[scale=0.55]{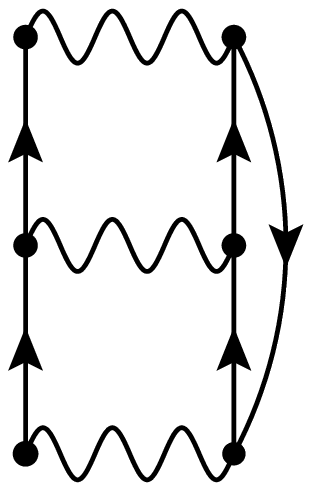}}
  \hspace{2cm}
    \subfloat[]{\label{3ord_b}\includegraphics[scale=0.55]{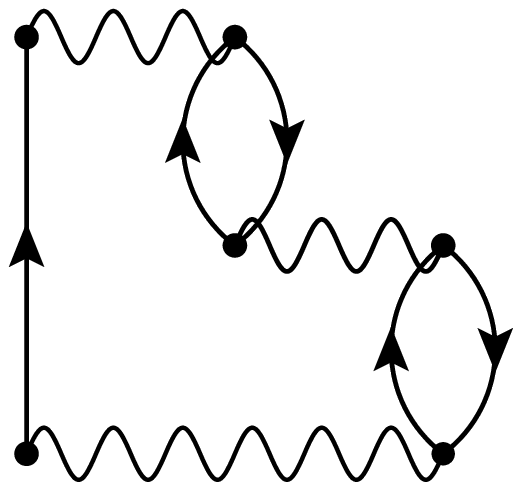}}
    \hspace{2cm}
    \subfloat[]{\label{3ord_c}\includegraphics[scale=0.55]{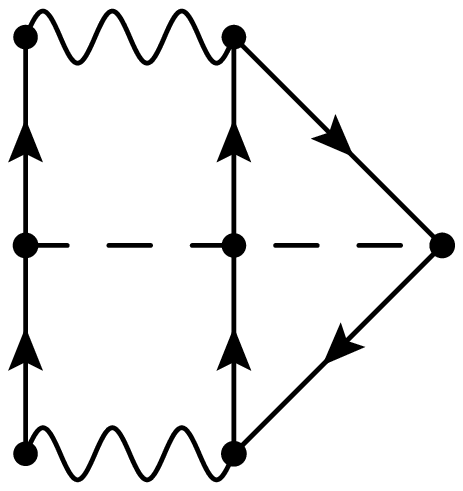}}
  \caption{ Three one-particle irreducible, skeleton and interaction-irreducible self-energy diagrams appearing at third order in the perturbative expansion of $\widetilde{\Sigma}(\omega) $. Only the third-order diagrams with at most $2p1h$ and $2h1p$ intermediate states are shown.}
  \label{3ord_a_b_c}
\end{figure}

The perturbative expansion of the self-energy generates 17 interaction-irreducible Feynman diagrams at third order~\cite{CarAr13}. Of these, only the three shown in Fig.~\ref{3ord_a_b_c} propagate at most $2p1h$ and $2h1p$ ISCs, whereas the remaining diagrams (not shown here) entail  at least  some contribution from $3p2h$ or $3h2p$ configurations. Moreover, Figs.~\ref{3ord_a} and~\ref{3ord_b} are the sole diagrams that do not involve any interaction-irreducible 3N term.  Given that for most systems three-body forces are weaker than the corresponding two-body ones~\footnote{For nuclear physics, one  may estimate that $<\hat{W}>\approx\frac1{10}<\hat{V}>$~\cite{grange1989,Epel09}.}, we expect that the  contributions of Fig.~\ref{3ord_a} and~\ref{3ord_b} are the most important and that diagram~\ref{3ord_c} is next in order of relevance, while the remaining 14 diagrams will not be dominant.
In this section, we present the explicit expressions of the coupling and interaction matrices entering in the ADC(3) formalism, at third order, as derived from the three aforementioned diagrams.

As for the second order, a set of expressions for the matrices $\textbf{M}$, $\textbf{N}$, $\textbf{C}$, $\textbf{D}$ and $E$ at ADC(3) is obtained from the direct comparison between the equations of the Feynman diagrams of Fig.~\ref{3ord_a_b_c}, with the general form of the self-energy in Eq.~(\ref{irr_SE_EXPA}).
The different terms can be organised according to the kind of interactions appearing in their contribution. For instance,  the diagrams~\ref{3ord_a} and~\ref{3ord_b} give contributions to the coupling matrices $\textbf{M}^{(\textrm{II})}$ and $\textbf{N}^{(\textrm{II})}$ that involve  two effective 2N  interactions and will be labelled with a \hbox{``$\textrm{2N} \; \textrm{2N}$''} superscript. 
 
\begin{figure}[t]
  \centering
    \subfloat[]{\label{M_2N_2N_nnk_a}\includegraphics[scale=0.4]{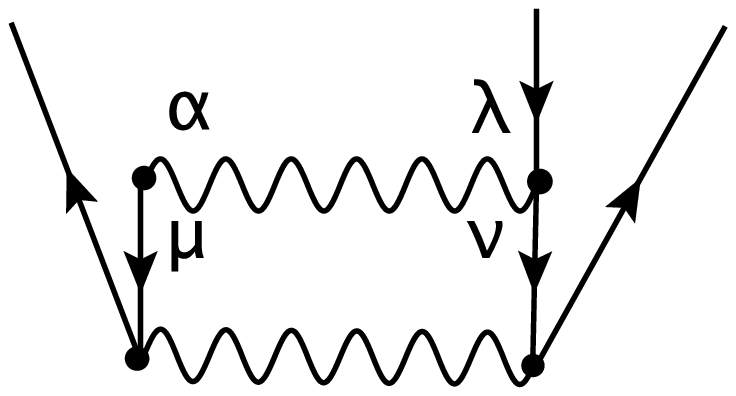}}
  \hspace{0.1cm}
    \subfloat[]{\label{M_2N_2N_nnk_b}\includegraphics[scale=0.35]{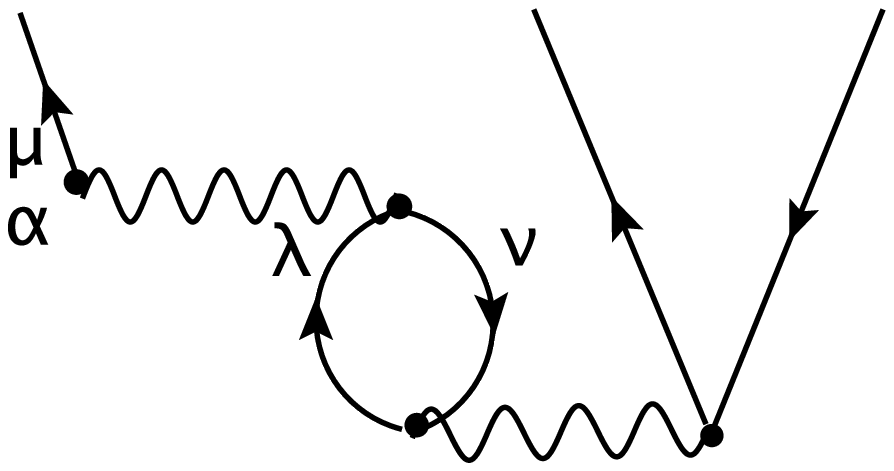}}
    \\
    \vspace{-2.cm}
    \subfloat[]{\label{N_2N_2N_kkn_a}\includegraphics[scale=0.4]{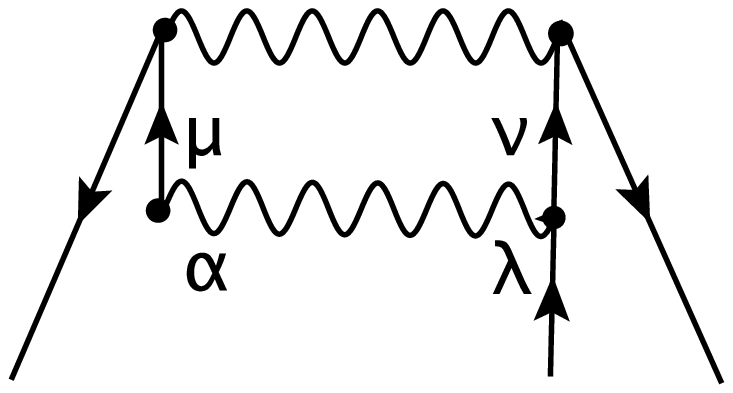}}
  \hspace{0.1cm}
    \subfloat[]{\label{N_2N_2N_kkn_b}\includegraphics[scale=0.35]{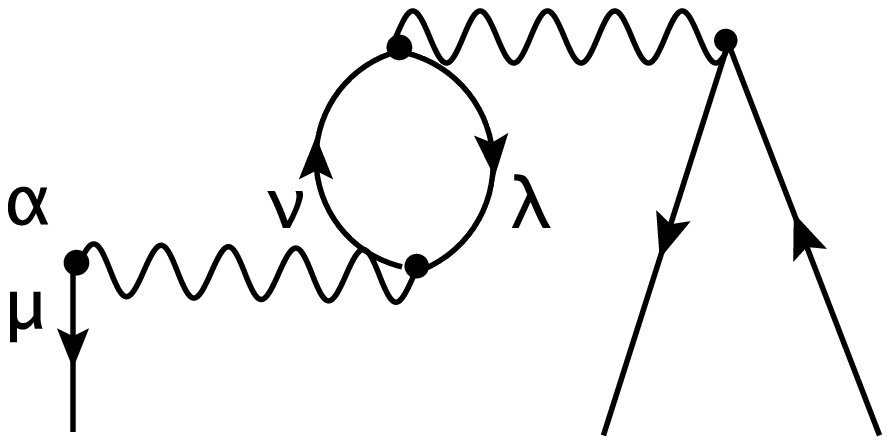}}
  \caption{ Diagrams of the ADC(3) coupling matrices with two effective 2NFs $\widetilde{V}$ that link to $2p1h$ ISCs (first row) and $2h1p$ ISCs (second row). The coupling matrices (a) and (b) correspond to Eqs.~(\ref{eq:M_3a},\ref{eq:M_3b}), while the coupling matrices (c) and (d) to Eqs.~(\ref{eq:N_3a},\ref{eq:N_3b}).}
  \label{M_N_3order_2N_2N}
\end{figure}

 Figures~\ref{3ord_a} and~\ref{3ord_b} generate all order summations of ladder and ring diagrams, respectively. These contain {\em only} effective 2NFs and their ADC(3) equations are well known~\cite{Schirmer1983,Carlo2016}. Hence, we simply states the results here.
 The forward-in-time coupling matrix arising from Fig.~\ref{3ord_a} is given by
\begin{eqnarray}
\label{eq:M_3a}
\textbf{M}^{(\textrm{2N 2N a})}_{ (n_1 n_2 k_3) \alpha} \equiv \frac{1}{2\sqrt{2}}  \,  \frac{ \cX_{\rho}^{n_1}  \cX_{\sigma}^{n_2} \ \widetilde{V}_{\rho \sigma ,\gamma \delta} \ \cY_{\gamma}^{k_4} \cY_{\delta}^{k_5}    }{\varepsilon_{k_4}^{-}+\varepsilon_{k_5}^{-}-\varepsilon_{n_1}^{+}-\varepsilon_{n_2}^{+}}\ (\cY_{\mu}^{k_4} \cY_{\nu}^{k_5})^*\cY_{\lambda}^{k_3} \ \widetilde{V}_{\mu\nu,\alpha \lambda} \, . 
\end{eqnarray}
The ring diagram of Fig.~\ref{3ord_b} gives rise to the forward-in-time coupling matrix,
\begin{eqnarray}
\label{eq:M_3b}
\textbf{M}^{(\textrm{2N 2N b})}_{(n_1 n_2 k_3) \alpha } &\equiv & \frac{1}{\sqrt{2}}   \left( \frac{   \cX_{\sigma}^{n_2} \cY_{\delta}^{k_3} \ \widetilde{V}_{ \sigma  \rho, \delta \gamma } \  \cY_{\gamma}^{k_5}  \cX_{\rho}^{n_4}  }{\varepsilon_{k_3}^{-}-\varepsilon_{n_2}^{+}+\varepsilon_{k_5}^{-}-\varepsilon_{n_4}^{+}}\ \cX_{\mu}^{n_1} (\cY_{\nu}^{k_5}   \cX_{\lambda}^{n_4})^* \ \widetilde{V}_{\mu\nu,\alpha \lambda} \right. \nonumber \\
&& \left.- \frac{  \cX_{\sigma}^{n_1} \cY_{\delta}^{k_3}  \ \widetilde{V}_{ \sigma \rho, \delta \gamma } \ \cY_{\gamma}^{k_5}  \cX_{\rho}^{n_4}   }{\varepsilon_{k_3}^{-}-\varepsilon_{n_1}^{+}+\varepsilon_{k_5}^{-}-\varepsilon_{n_4}^{+}} \ \cX_{\mu}^{n_2} (\cY_{\nu}^{k_5}   \cX_{\lambda}^{n_4})^* \ \widetilde{V}_{\mu\nu,\alpha \lambda} \right)\, , 
\end{eqnarray}
which is explicitly antisymmetrized with respect to the $n_1$ and $n_2$ fermion lines. The diagrammatic representations of the two coupling matrices of Eqs.~(\ref{eq:M_3a}) and~(\ref{eq:M_3b}) are depicted in Figs.~\ref{M_2N_2N_nnk_a} and~\ref{M_2N_2N_nnk_b}, respectively.

For the same self-energy diagrams of Figs.~\ref{3ord_a} and~\ref{3ord_b} but from the backward-in-time Goldstone diagrams, we find the coupling matrices
\begin{eqnarray}
\label{eq:N_3a}
\textbf{N}^{(\textrm{2N 2N a})}_{\alpha (k_1 k_2 n_3)} \equiv \frac{1}{2\sqrt{2}} \, \widetilde{V}_{\alpha \lambda,\mu\nu} \ \cX_{\lambda}^{n_3}  (\cX_{\mu}^{n_4} \cX_{\nu}^{n_5} )^* \ \frac{ \cX_{\rho}^{n_4}  \cX_{\sigma}^{n_5}  \ \widetilde{V}_{\rho \sigma,\gamma \delta} \ \cY_{\gamma}^{k_1}   \cY_{\delta}^{k_2} }{\varepsilon_{k_1}^{-}+\varepsilon_{k_2}^{-}-\varepsilon_{n_4}^{+}-\varepsilon_{n_5}^{+}}
\end{eqnarray}
and
\begin{eqnarray}
\label{eq:N_3b}
\textbf{N}^{(\textrm{2N 2N b})}_{\alpha (k_1 k_2 n_3)} &\equiv & \frac{1}{\sqrt{2}}   \left( \widetilde{V}_{\alpha \lambda, \mu\nu} \ (\cY_{\lambda}^{k_5})^*  \cY_{\mu}^{k_1} (\cX_{\nu}^{n_4})^*     \    \frac{  \cX_{\sigma}^{n_4} \cY_{\delta}^{k_5}   \ \widetilde{V}_{\sigma \rho , \delta \gamma } \  \cY_{\gamma}^{k_2}  \cX_{\rho}^{n_3}  }{\varepsilon_{k_2}^{-}-\varepsilon_{n_3}^{+}+\varepsilon_{k_5}^{-}-\varepsilon_{n_4}^{+}} \ \right. \nonumber \\
&& \left.-  \widetilde{V}_{\alpha \lambda, \mu\nu} \ (\cY_{\lambda}^{k_5})^*  \cY_{\mu}^{k_2} (\cX_{\nu}^{n_4})^*     \ \frac{  \cX_{\sigma}^{n_4} \cY_{\delta}^{k_5} \ \widetilde{V}_{ \sigma \rho, \delta \gamma} \ \cY_{\gamma}^{k_1}  \cX_{\rho}^{n_3}   }{\varepsilon_{k_1}^{-}-\varepsilon_{n_3}^{+}+\varepsilon_{k_5}^{-}-\varepsilon_{n_4}^{+}}\right)\; . 
\end{eqnarray}
Their diagrammatic representation is displayed in Figs.~\ref{N_2N_2N_kkn_a} and~\ref{N_2N_2N_kkn_b} respectively, where it is clear that they are linked to the $2h1p$ propagators.

The interaction matrices are found by comparing the third order Goldstone diagrams with double poles to the third and sixth lines of~Eq.~\eqref{irr_SE_EXPA}.
 The interaction matrix connecting $2p1h$ propagators through a particle-particle (pp), ladder, interaction is
 \begin{equation}
 \label{int_term_5a_C}
 \textbf{C}^{pp}_{\substack{(n_1 n_2 k_3), (n_4 n_5 k_6)}} \equiv \frac{1}{2} \, \cX_{\mu}^{n_1}   \cX_{\nu}^{n_2} \,  \widetilde{V}_{\mu\nu, \lambda \rho} \,  (\cX_{\lambda}^{n_4}  \cX_{\rho}^{n_5})^* \,  \delta_{k_3 k_6} \; ,
 \end{equation}
 while the one connecting through particle-hole (ph) rings is composed by four terms that arise from the antisymmetrization with respect to the $n_1$ and $n_2$ particles to the left and the $n_4$ and $n_5$ ones to the right,
 \begin{eqnarray}
 \label{int_term_5b_C}
  \textbf{C}^{ph}_{\substack{(n_1 n_2 k_3), (n_4 n_5 k_6)}} & = &  \frac{1}{2}\left(\cX_{\nu}^{n_2} \cY_{\rho}^{k_3}  \, \widetilde{V}_{\mu\nu, \lambda \rho} \, (\cX_{\lambda}^{n_5}   \cY_{\mu}^{k_6})^* \, \delta_{n_1 n_4}   \right. \nonumber \\ 
  && \left.  - \, \cX_{\nu}^{n_2} \cY_{\rho}^{k_3} \, \widetilde{V}_{\mu\nu, \lambda \rho} \, (\cX_{\lambda}^{n_4} \cY_{\mu}^{k_6})^*   \, \delta_{n_1 n_5} \right. \nonumber \\ 
  && \left. - \, \cX_{\nu}^{n_1}   \cY_{\rho}^{k_3} \, \widetilde{V}_{\mu\nu, \lambda \rho} \, (\cX_{\lambda}^{n_5}\cY_{\mu}^{k_6})^*  \, \delta_{n_2 n_4} \right. \nonumber \\ 
  && \left. + \, \cX_{\nu}^{n_1}  \cY_{\rho}^{k_3} \,  \widetilde{V}_{\mu\nu, \lambda \rho} \,  (\cX_{\lambda}^{n_4} \cY_{\mu}^{k_6})^*   \, \delta_{n_2 n_5}\right) \, .
 \end{eqnarray}

%
In the backward-in-time self-energy Goldstone diagrams,  the interaction matrices  connecting $2h1p$ propagators through a hole-hole (hh) interaction lead to
 \begin{equation}
 \label{int_term_5a_D}
 \textbf{D}^{hh}_{\substack{(k_1 k_2 n_3), (k_4 k_5 n_6)}} \equiv -\frac{1}{2} \,  (\cY_{\mu}^{k_1}   \cY_{\nu}^{k_2})^* \, \widetilde{V}_{\mu\nu, \lambda \rho} \, \cY_{\lambda}^{k_4}  \cY_{\rho}^{k_5} \, \delta_{n_3 n_6} \; ,
 \end{equation}
 while the one connecting through a hole-particle (hp) interaction gives
 \begin{eqnarray}
 \label{int_term_5b_D}
  \textbf{D}^{hp}_{\substack{(k_1 k_2 n_3), (k_4 k_5 n_6)}} & = & \frac{1}{2} \left(      (\cY_{\mu}^{k_2} \cX_{\rho}^{n_3})^* \, \widetilde{V}_{\mu\nu, \lambda \rho} \, \cY_{\lambda}^{k_5} \cX_{\nu}^{n_6} \,   \delta_{k_1 k_4}  \right. \nonumber \\ 
  && \left.  - \,
  (\cY_{\mu}^{k_2} \cX_{\rho}^{n_3})^* \, \widetilde{V}_{\mu\nu, \lambda \rho} \, \cY_{\lambda}^{k_4}    \cX_{\nu}^{n_6}  \,  \delta_{k_1 k_5}  \right. \nonumber \\ 
  && \left.  - \,
   (\cY_{\mu}^{k_1} \cX_{\rho}^{n_3})^* \, \widetilde{V}_{\mu\nu, \lambda \rho} \, \cY_{\lambda}^{k_5}  \cX_{\nu}^{n_6} \,  \delta_{k_2 k_4} \right. \nonumber \\ 
  && \left. +   \,  (\cY_{\mu}^{k_1} \cX_{\rho}^{n_3})^* \, \widetilde{V}_{\mu\nu, \lambda \rho} \, \cY_{\lambda}^{k_4}  \cX_{\nu}^{n_6} \, \delta_{k_2 k_5} 
\right) \; .
 \end{eqnarray}


We now turn to the Feynman diagram of Fig.~\ref{3ord_c}, which is the focus of the present work. To our knowledge the ADC formulas arising from this term have not been presented before. The Feynman rules give the following expression for it:
\begin{eqnarray}
\label{Fey_2c}
\Sigma^{(3c)}_{\alpha \beta}(\omega)&=&
\! -\frac{( \hbar)^{4} }{4}
\! \int \! \frac{{\mathrm d}\omega_1}{2\pi \mathrm{i}}\! \int \! \frac{{\mathrm d}\omega_2}{2\pi \mathrm{i}} \! \int \! \frac{{\mathrm d}\omega_3}{2\pi \mathrm{i}} \! \int \! \frac{{\mathrm d}\omega_4}{2\pi \mathrm{i}}
\! \!  \sum_{\substack{ \gamma\delta\nu \mu\epsilon\lambda \\ \xi\eta\theta \sigma \tau\chi}} \! \!
\widetilde{V}_{\alpha\gamma,\delta\nu} \ g_{\xi \gamma}(\omega_3) \ g_{\nu \lambda}(\omega -\omega_1+\omega_3)  \nonumber \\ &&
 g_{\delta \epsilon}(\omega_1)\ W_{\mu\epsilon\lambda,\xi\eta\theta} \ g_{\theta \tau}(\omega-\omega_2+\omega_4) \ g_{\eta \sigma}(\omega_2) \ g_{\chi \mu}(\omega_4) \ \widetilde{V}_{\sigma\tau,\beta\chi} \; . 
\end{eqnarray}
By performing the four integrals  in the complex plane, we find six terms, corresponding to the different time orderings of the three interactions. Altogether we obtain,
\begin{align}
\label{2c_expr}
\Sigma^{(3c)}_{\alpha \beta}(\omega)&=
 \frac{1}{4}
\sum_{\substack{ \gamma\delta\nu \mu\epsilon\lambda \\ \xi\eta\theta \sigma \tau\chi}}  \widetilde{V}_{\alpha\gamma,\delta\nu}
W_{\epsilon\lambda\mu, \eta\theta\xi} \widetilde{V}_{\sigma\tau,\beta\chi}\times \nonumber \\
&\left(  -\sum_{\substack{ n_1 n_2 k_3 \\ n_4 n_5 k_6 }}  \frac{(\cX_{\delta}^{n_1} \cX_{\nu}^{n_2} \cY_{\gamma}^{k_3})^*    \cX_{\epsilon}^{n_1}  \cX_{\lambda}^{n_2} \cY_{\xi}^{k_3}    (\cX_{\eta}^{n_4}\cX_{\theta}^{n_5}\cY_{\mu}^{k_6}  )^* \cX_{\sigma}^{n_4} \cX_{\tau}^{n_5} \cY_{\chi}^{k_6}  }{\left(\hbar \omega - ( \varepsilon_{n_1}^{+}+\varepsilon_{n_2}^{+}-\varepsilon_{k_3}^{-})+ \textrm{i} \eta\right) \left(\hbar \omega - (\varepsilon_{n_4}^{+}+\varepsilon_{n_5}^{+}-\varepsilon_{k_6}^{-} ) + \textrm{i} \eta\right) } 
 \right. \nonumber \\
&\left.  +\sum_{\substack{ k_1 k_2 n_3 \\ n_4 n_5 k_6 }} \frac{ \cY_{\delta}^{k_1} \cY_{\nu}^{k_2} \cX_{\gamma}^{n_3}  (\cY_{\epsilon}^{k_1} \cY_{\lambda}^{k_2} \cY_{\mu}^{k_6} \cX_{\eta}^{n_4}\cX_{\theta}^{n_5} \cX_{\xi}^{n_3})^*  \cX_{\sigma}^{n_4} \cX_{\tau}^{n_5}   \cY_{\chi}^{k_6}  }{ \left(\varepsilon_{k_1}^{-}+\varepsilon_{k_2}^{-} +\varepsilon_{k_6}^{-}-\varepsilon_{n_3}^{+}-\varepsilon_{n_4}^{+}-\varepsilon_{n_5}^{+}\right)\left(\hbar \omega - \left(\varepsilon_{n_4}^{+}+\varepsilon_{n_5}^{+}-\varepsilon_{k_6}^{+}\right) + \textrm{i} \eta\right) }  \right. \nonumber \\
&\left.  +\sum_{\substack{ k_1 k_2 n_3 \\ n_4 n_5 k_6 }} \frac{(\cX_{\delta}^{n_4}\cX_{\nu}^{n_5}\cY_{\gamma}^{k_6})^*   \cX_{\epsilon}^{n_4}  \cX_{\lambda}^{n_5} \cX_{\mu}^{n_3}     \cY_{\eta}^{k_1}  \cY_{\theta}^{k_2}  \cY_{\xi}^{k_6}  (\cY_{\sigma}^{k_1}\cY_{\tau}^{k_2}\cX_{\chi}^{n_3})^*   }{\left(\hbar \omega - \left(\varepsilon_{n_4}^{+}+\varepsilon_{n_5}^{+}-\varepsilon_{k_6}^{-} \right) + \textrm{i} \eta\right)\left(\varepsilon_{k_1}^{-}+\varepsilon_{k_2}^{-} +\varepsilon_{k_6}^{-}-\varepsilon_{n_3}^{+}-\varepsilon_{n_4}^{+}-\varepsilon_{n_5}^{+}\right) }  \right. \nonumber \\
& \left.  -   \sum_{\substack{ k_1 k_2 n_3 \\ k_4 k_5 n_6 }}  \frac{\cY_{\delta}^{k_1} \cY_{\nu}^{k_2}   \cX_{\gamma}^{n_3}   (\cY_{\epsilon}^{k_1} \cY_{\lambda}^{k_2}\cX_{\xi}^{n_3})^*    \cY_{\eta}^{k_4} \cY_{\theta}^{k_5} \cX_{\mu}^{n_6}  (\cY_{\sigma}^{k_4}\cY_{\tau}^{k_5}\cX_{\chi}^{n_6})^*  }{\left(\hbar \omega - ( \varepsilon_{k_1}^{-}+\varepsilon_{k_2}^{-}-\varepsilon_{n_3}^{+})-\textrm{i} \eta\right) \left(\hbar \omega - (\varepsilon_{k_4}^{-}+\varepsilon_{k_5}^{-}-\varepsilon_{n_6}^{+} )-\textrm{i} \eta\right) } \right. \nonumber \\
&\left.  -\sum_{\substack{ k_1 k_2 n_3 \\ n_4 n_5 k_6 }} \frac{   \cY_{\delta}^{k_1} \cY_{\nu}^{k_2}  \cX_{\gamma}^{n_3}   (\cY_{\epsilon}^{k_1} \cY_{\lambda}^{k_2} \cY_{\mu}^{k_6} \cX_{\eta}^{n_4}\cX_{\theta}^{n_5} \cX_{\xi}^{n_3})^* \cX_{\sigma}^{n_4} \cX_{\tau}^{n_5} \cY_{\chi}^{k_6} }{\left(\hbar \omega - \left(\varepsilon_{k_1}^{-}+\varepsilon_{k_2}^{-}-\varepsilon_{n_3}^{+} \right)-\textrm{i} \eta\right) \left(\varepsilon_{k_1}^{-}+\varepsilon_{k_2}^{-} +\varepsilon_{k_6}^{-}-\varepsilon_{n_3}^{+}-\varepsilon_{n_4}^{+}-\varepsilon_{n_5}^{+}\right) }  \right. \nonumber \\
&\left. -\sum_{\substack{ k_1 k_2 n_3 \\ n_4 n_5 k_6 }} \frac{  (\cX_{\delta}^{n_4}\cX_{\nu}^{n_5}\cY_{\gamma}^{k_6})^*      \cX_{\epsilon}^{n_4} \cX_{\lambda}^{n_5} \cX_{\mu}^{n_3}   \cY_{\eta}^{k_1} \cY_{\theta}^{k_2} \cY_{\xi}^{k_6}  (\cY_{\sigma}^{k_1} \cY_{\tau}^{k_2} \cX_{\chi}^{n_3})^*}{\left( \varepsilon_{k_1}^{-}+\varepsilon_{k_2}^{-} +\varepsilon_{k_6}^{-}-\varepsilon_{n_3}^{+}-\varepsilon_{n_4}^{+}-\varepsilon_{n_5}^{+} \right)\left(\hbar \omega - \left(\varepsilon_{k_1}^{-}+\varepsilon_{k_2}^{-}-\varepsilon_{n_3}^{-} \right) -\textrm{i} \eta\right)  }   \right) \, ,
\end{align}
where the first (last) three terms correspond to forward-in-time (backward-in-time) Goldstone diagrams.

 \begin{figure}[t]
  \centering
    \subfloat[]{\label{M_2N_3N_3c}\includegraphics[scale=0.37]{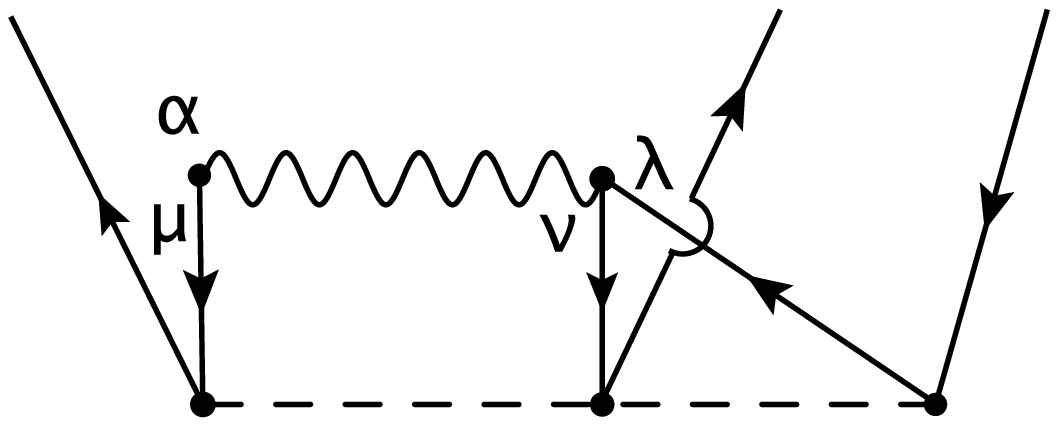}}
  \hspace{0.1cm}
    \subfloat[]{\label{N_2N_3N_3c}\includegraphics[scale=0.37]{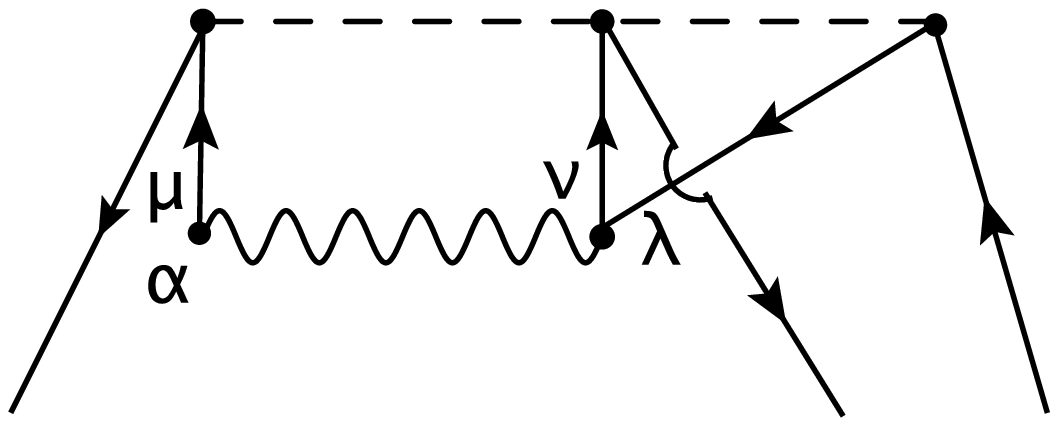}}
  \caption{ Diagrams of the ADC(3) coupling matrices  with one effective 2NF $\widetilde{V}$  and one interaction-irreducible 3NF $\hat{W}$. The coupling matrix (a) is linked to $2p1h$ ISCs and corresponds to Eq.~(\ref{eq:M_3c}),
while  (b) is linked to $2h1p$ ISCs and corresponds to Eq.~(\ref{eq:N_3c}).}
  \label{M_N_3order_2N_3N_3c}
\end{figure} 
By comparing to the third order terms in Eq.~\eqref{irr_SE_EXPA}, one see that the new contributions to the coupling matrices contain one effective 2NF and one interaction-irreducible 3NF. The following forward-in-time matrix can be singled out from either  the second or third line of Eq.~(\ref{2c_expr}),
\begin{eqnarray}
\label{eq:M_3c}
\textbf{M}^{(\textrm{2N 3N a})}_{(n_1 n_2 k_3) \alpha } \equiv \frac{1}{2\sqrt{2}} \,  \frac{ \cX_{\xi}^{n_4} \cX_{ \rho}^{n_1}  \cX_{\sigma}^{n_2}    \ {W}_{\xi \rho \sigma , \zeta \eta \theta} \ \cY_{\zeta}^{k_3}   \cY_{\eta}^{k_5}  \cY_{\theta}^{k_6} }{\varepsilon_{k_3}^{-}+\varepsilon_{k_5}^{-}+\varepsilon_{k_6}^{-}-\varepsilon_{n_1}^{+}-\varepsilon_{n_2}^{+}-\varepsilon_{n_4}^{+}}  \ (\cY_{\mu}^{k_5} \cY_{\nu}^{k_6}   \cX_{\lambda}^{n_4})^* \ \widetilde{V}_{\mu\nu,\alpha \lambda} \; ,
\end{eqnarray}
while in the last two lines of Eq.~(\ref{2c_expr}) we read the backward-in-time coupling matrix:
\begin{eqnarray}
\label{eq:N_3c}
\textbf{N}^{(\textrm{2N 3N a})}_{\alpha (k_1 k_2 n_3)} \equiv - \frac{1}{2\sqrt{2}} \, \widetilde{V}_{\alpha \lambda,\mu\nu} \ (\cY_{\lambda}^{k_4} \cX_{\mu}^{n_5}   \cX_{\nu}^{n_6} )^* \frac{ \cX_{\rho}^{n_3}  \cX_{\sigma}^{n_5}  \cX_{\xi}^{n_6}  \ {W}_{\rho \sigma \xi, \theta \zeta \eta } \ \cY_{\theta}^{k_4} \cY_{\zeta}^{k_1}   \cY_{\eta}^{k_2}  }{\varepsilon_{k_1}^{-}+\varepsilon_{k_2}^{-}+\varepsilon_{k_4}^{-}-\varepsilon_{n_3}^{+}-\varepsilon_{n_5}^{+}-\varepsilon_{n_6}^{+}} \; .
\end{eqnarray}
The diagrammatic representations of Eqs.~\eqref{eq:M_3c} and~\eqref{eq:N_3c} are displayed in Fig.~\ref{M_N_3order_2N_3N_3c}.
 
  The only interaction matrix that connects $2p1h$ ISCs through a 3NF is found from the first term of Eq.~(\ref{2c_expr}),
  \begin{eqnarray}
 \label{int_term_5c_C}
 \textbf{C}^{3N}_{\substack{(n_1 n_2 k_3), (n_4 n_5 k_6)}} \equiv  -\frac{1}{2} \, \cX_{\nu}^{n_1}   \cX_{\mu}^{n_2} \cY_{\rho}^{k_3}  \, W_{ \nu\mu \lambda,   \epsilon \eta \rho} \, (\cX_{\epsilon}^{n_4}  \cX_{\eta}^{n_5} \cY_{\lambda}^{k_6})^*  \; , 
 \end{eqnarray}
which is explicitly antisymmetric in the particle indexes. With Eqs.~(\ref{int_term_5c_C}) and~(\ref{eq:M_2a}) we can rewrite the first term of Eq.~(\ref{2c_expr}) as,
 \begin{eqnarray}
 \label{2c_expr_I}
\textbf{M}^{\dagger (\textrm{I-2N})}_{\alpha r} \, \frac{1}{\hbar \omega - E_r} \, \textbf{C}^{3N}_{r r'} \, \frac{1}{\hbar \omega - E_{r'}} \, \textbf{M}^{(\textrm{I-2N})}_{r' \beta} \; .
 \end{eqnarray}
The expression~(\ref{2c_expr_I}) contains only the first order contribution in the interaction matrix expansion, corresponding to the second term in the r.h.s. of Eq.~(\ref{geo_Serie}), for $B = \textbf{C}^{3N}$. This is resummed to all order by diagonalizing the Dyson matrix~(\ref{eq:Dy_residual_5}),  which will automatically include \emph{all} the higher order terms in the expansion.
 
From the fourth term of Eq.~(\ref{2c_expr}), we single out the only backward-in-time interaction matrix connecting two $2h1p$ configurations through a 3N interaction, that is
\begin{eqnarray}
 \label{int_term_5c_D}
 \textbf{D}^{3N}_{\substack{(k_1 k_2 n_3), (k_4 k_5 n_6)}} \equiv  -\frac{1}{2} \, (\cY_{\nu}^{k_1}   \cY_{\mu}^{k_2} \cX_{\rho}^{n_3})^* \, W_{ \nu\mu \lambda,  \epsilon \eta  \rho} \,\cY_{\epsilon}^{k_4}  \cY_{\eta}^{k_5}  \cX_{\lambda}^{n_6}   \; , 
\end{eqnarray}
which is also explicitly antisymmetric in the hole indexes. With Eqs.~(\ref{int_term_5c_D}) and~(\ref{eq:N_2a}) we associate the fourth term of Eq.~(\ref{2c_expr}) to
 \begin{eqnarray}
 \label{2c_expr_IV}
\textbf{N}^{(\textrm{I-2N})}_{\alpha s} \, \frac{1}{\hbar \omega - E_s} \, \textbf{D}^{3N}_{s s'} \, \frac{1}{\hbar \omega - E_{s'}} \, \textbf{N}^{\dagger(\textrm{I-2N})}_{s' \beta} \; .
 \end{eqnarray}
We stress again the fact that Eq.~(\ref{2c_expr_IV}), being a first-order term in $\textbf{D}^{3N}$, is resummed with \emph{all} the other higher order contributions when solving the Dyson equation.

Finally, the ADC(3) working equations for the set of Feynman diagrams in Fig.~\ref{3ord_a_b_c} is summarized by the following expressions,
\begin{eqnarray}
\label{ADC3_somm_I}
\textbf{M}^{(\textrm{II})}_{j \alpha}  &= & \textbf{M}^{(\textrm{2N 2N a})}_{ (n_1 n_2 k_3) \alpha}  \quad +   \qquad \textbf{M}^{(\textrm{2N 2N b})}_{ (n_1 n_2 k_3) \alpha} \quad +  \qquad\textbf{M}^{(\textrm{2N 3N a})}_{ (n_1 n_2 k_3) \alpha} \,, \\
\label{ADC3_somm_II}\textbf{N}^{(\textrm{II})}_{\alpha k}  & = &  \textbf{N}^{(\textrm{2N 2N a})}_{\alpha (k_1 k_2 n_3)} \quad +   \qquad \textbf{N}^{(\textrm{2N 2N b})}_{\alpha (k_1 k_2 n_3)}  \quad +   \qquad \textbf{N}^{(\textrm{2N 3N a})}_{\alpha (k_1 k_2 n_3)} \, ,\\
\label{somm_C_3}\textbf{C}_{j j'}  & = &   \textbf{C}^{pp}_{\substack{(n_1 n_2 k_3), (n_4 n_5 k_6)}} +   \textbf{C}^{ph}_{\substack{(n_1 n_2 k_3), (n_4 n_5 k_6)}} +  \textbf{C}^{3N}_{\substack{(n_1 n_2 k_3), (n_4 n_5 k_6)}}  \, , \\
\label{somm_D_3}\textbf{D}_{k k'}  & =  & \textbf{D}^{hh}_{\substack{(k_1 k_2 n_3), (k_4 k_5 n_6)}}  +   \textbf{D}^{hp}_{\substack{(k_1 k_2 n_3), (k_4 k_5 n_6)}} +  \textbf{D}^{3N}_{\substack{(k_1 k_2 n_3), (k_4 k_5 n_6)}}  \, .
\end{eqnarray}

At third order, beside the equations summarized above, there are the coupling and interaction matrices imposed by the remaining 14 one-particle irreducible, skeleton and interaction-irreducible self-energy diagrams, which are topologically distinct from Fig.~\ref{3ord_a_b_c}~\cite{CarAr13}. The expressions of the coupling and interaction matrices derived from all these diagrams contribute to the $3p2h$ and $3h2p$ sectors of Eq.~(\ref{eq:Dy_residual_5})~\cite{Raimondi_forth}.

\section{\label{end}Summary and outlook}
We have shown the working equations for the ADC(3) formalism applied to the irreducible self-energy in the SCGF formalism, with 2NFs and 3NFs.
This formalism allows an efficient and accurate numerical implementation for the solution of the Dyson equation, which is recast as an energy eigenvalue problem. Moreover, within a given order, the matrix form of  the Dyson equation allows the infinite resummation of certain classes of diagrams, specifically ladder and ring diagrams, preserving the non-perturbative nature of the SCGF approach.

The minimal expressions for both the coupling and interaction matrices, required to conform to the structure of the self-energy as analytic function in the energy plane, have been revisited completely at the  ADC(2) level. We have displayed the most important terms at third order, and we derived for the first time  the coupling and interaction matrices of the Feynman diagram in Fig.~\ref{3ord_c}, containing one interaction-irreducible 3NF. This term is relevant because it is the only irreducible 3NF insertion that links to the dominant ISCs in the self-energy, that is $2p1h$ and $2h1p$ intermediate excitations.  The complete ADC(3) working equations for the Dyson SCGF approach will be presented in a forthcoming publication~\cite{Raimondi_forth}, while the extension  of the Gorkov SCGF formalism of Ref.~\cite{Soma2011} to ADC(3) is part of  future plans.

\bibliography{Procs_NTSE2016_VI}
\bibliographystyle{ieeetr}

\end{document}